\documentstyle[11pt,moriond,epsfig]{article}
\def\Journal#1#2#3#4{{#1} {\bf #2}, #3 (#4)}

\def\PRB{{\em Phys. Rev.} B}

\def\be{\begin{equation}}
\def\ee{\end{equation}}
\def\bea{\begin{eqnarray}}
\def\eea{\end{eqnarray}}

\begin{document}
%\vspace*{1cm}
\title{FRACTIONAL-QUANTUM-HALL EDGES AT FILLING FACTOR $\mathbf \nu=1-1/m$}

\author{ U.~Z\"ULICKE}

\address{Institut f\"ur Theoretische Festk\"orperphysik, Universit\"at
Karlsruhe, D-76128 Karlsruhe, Germany}

\author{ A.~H.~MacDonald }

\address{Department of Physics, Indiana University, Bloomington, Indiana
47405, U.S.A.}

\author{ M.~D.~Johnson }

\address{Department of Physics, University of Central Florida, Orlando,
Florida 32816, U.S.A.}

\maketitle\abstracts{
We consider the edge of a two-dimensional electron system that is in the
quantum-Hall-effect regime at filling factor $\nu=1-1/m$ with $m$ being an odd
integer, where microscopic theory explaining the occurrence of the quantum
Hall effect in the bulk predicts the existence of two counterpropagating
edge-excitation modes. These two modes are the classical edge-magnetoplasmon
mode and a slow-moving neutral mode. Assuming the electrons to be confined by
a coplanar neutralizing background of positive charges, and taking careful
account of long-range Coulomb interactions, we determine microscopically the
velocity $v_{\mathrm n}$ of the neutral mode and the edge width $d$. Our
results are intended to guide experimental efforts aimed at verifying the
existence of the neutral mode, which would provide a powerful confirmation of
the current microscopic understanding of quantum-Hall physics at the simplest
hierarchical filling factors $\nu=1-1/m$.
}

\section{Introduction}

The quantum Hall (QH) effect~\cite{qhe} occurs in two-dimensional
(2D) electron systems whenever an incompressibility develops~\cite{ahmintro} in
the bulk at a magnetic-field($B$)-dependent value of the electronic sheet
density $n_{\mathrm e}$. Experimentally, the QH effect is observed when the
filling factor $\nu=2\pi\ell^2 n_{\mathrm e}$ is equal to an integer or certain
fractions. (Here we defined the magnetic length $\ell=\sqrt{\hbar c/|e B|}$.)
The physical origin of the incompressibility, i.e., a gap for excitation of
unbound particle-hole pairs, is quite different for the integer and fractional
QH effects. In the integer case, the incompressibility arises from Landau
quantization of the kinetic energy of a charged 2D particle moving in a
perpendicular magnetic field. At fractional filling factors where partially
filled Landau levels exist, incompressibility is a consequence of
electron-electron interactions. Exactly how interactions give rise to bulk
incompressibilities is most well-understood microscopically~\cite{rbl:prl:83}
for filling factors that are the inverse of an odd number, i.e., $\nu=1/m$ with
$m=3,5,\dots$. To explain incompressibility at other fractional values of the
filling factor where the QH effect is observed, hierarchical
models~\cite{hierarchy} have been proposed. We focus
here on QH systems at $\nu=1-1/m$, which can be regarded as the simplest
hierarchical filling factors.

In both the integer and fractional cases, the only low-lying excitations
present in a QH sample are localized at the boundary. In a magnetic field,
collective modes known as edge-magnetoplasmons~\cite{volkmikh} (EMP) occur at
the edge of a 2D electron system even when the bulk is compressible. Outside
of the QH regime, however, these modes have a finite life time due to decay
into incoherent particle-hole excitations and are most aptly described using a
hydrodynamic picture. In the QH regime, provided that the edge of the 2D
electron system is sufficiently sharp, the microscopic physics simplifies and
there is no particle-hole continuum into which the modes can decay.
Generalizations of models familiar from the study of one-dimensional (1D)
electron systems~\cite{voit:reprog:94} can then be used to provide a fully
microscopic description of integer~\cite{bih:prb:82} and
fractional~\cite{wen} QH
edges. In particular, the sharp edge of a QH system at $\nu=1/m$ supports a
single branch of chiral edge excitations. These are EMP modes which, in this
case, have an especially simple microscopic description. For the hierarchical
filling factors, however, both microscopic theory~\cite{ahm:prl:90} and
phenomenological considerations~\cite{wen} suggest that even a sharp
edge supports additional branches of chiral excitations, some of which can be
propagating in the direction opposite to the EMP mode. For $\nu=1-1/m$, a
single counterpropagating edge mode is expected to exist in addition to the
EMP mode. The prediction of this additional mode is entirely due to our
microscopic understanding of why the QH effect is observed at $\nu=1-1/m$;
such a prediction would not arise within a purely classical, hydrodynamic
theory for the interacting 2D electron system. Hence, experimental confirmation
of the additional counterpropagating edge mode would provide a strong 
confirmation for the predictive power of microscopic theory describing
fractional-QH physics. However, time-domain studies~\cite{ray:prb:92} of edge
modes at $\nu=2/3$ have turned up no evidence for this mode.

Here we review the microscopic model for a finite QH sample that is at a
filling factor $\nu=1-1/m$. We determine the two parameters characterizing the
edge, which are its stiffness to long-wave-length neutral excitations, and the
effective edge width. Performing a careful separation of long-range and
short-range pieces of interactions between edge excitations, we are able to
calculate the velocity of the counterpropagating mode.

\section{Microscopic Model for the Quantum-Hall Sample at $\mathbf\nu=1-1/m$}

\begin{figure}
%\rule{5cm}{0.2mm}\hfill\rule{5cm}{0.2mm}
%\vskip 2.5cm
%\rule{5cm}{0.2mm}\hfill\rule{5cm}{0.2mm}
\centerline{\psfig{figure=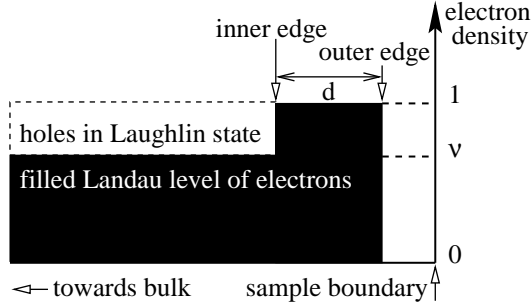,height=1.6in}}
\caption{Schematic profile of the electron density [in the direction
perpendicular to the edge, measured in units of $1/(2\pi\ell^2)$] of a QH
sample at filling factor $\nu=1-1/m$. Microscopic theory explains the
incompressibility at $\nu=1-1/m$ as due to incompressibility of a system
of {\em holes\/} at hole filling factor $1-\nu=1/m$. In a finite system, the
Laughlin state of holes has an edge where the electron density approaches
unity. Hence, there must be a second, outer edge of a filled lowest Landau
level of electrons. The local filling factor deviates from its uniform bulk
value near the edge.  In the abrupt edge model, this leads to a dipolar strip.
\label{phconj}}
\end{figure}
Laughlin~\cite{rbl:prl:83} proposed a set of many-electron states that are
incompressible at $\nu=1/m$ to explain the occurrence of the QH effect at
these filling factors. Hierarchical constructions relate the incompressibility
of the 2D electron system at $\nu=1/m$ to incompressibility at some other
filling factor. For example, due to the fact that the lowest Landau level
has perfect particle-hole symmetry,~\cite{smg:prb:84} in the absence of
Landau-level mixing a system of electrons at a fractional filling factor
$\nu_{\mathrm e}=\nu$ is equivalent to a {\em hole\/} system at filling factor
$\nu_{\mathrm h}=1-\nu$. Based on Laughlin's theory, the {\em hole\/} system
is incompressible at {\em hole\/} filling factors $\nu_{\mathrm h}=1/m$, which
translates into incompressibility of the corresponding electron system at
$\nu_{\mathrm e}=1-1/m$. This microscopic explanation of the occurrence of the
QH effect at $\nu=1-1/m$ implies that the ground state of the electron system
at that filling factor is the particle-hole conjugate of a Laughlin state for
holes at $\nu_{\mathrm h}=1/m$. In a finite electron system, there are then
{\em two\/} distinct ways to add particles at $\nu=1-1/m$; either before or
after performing the particle-hole-conjugation transformation. This leads us to
expect {\em two\/} distinct edges of the $\nu=1-1/m$ QH sample, an inner one
and an outer one, which are separated by a distance $d$. (See
Fig.~\ref{phconj}.) The inner edge is that of a Laughlin state of holes at
$\nu_{\mathrm h}=1/m$, and the outer one is the edge of a filled lowest Landau
level of electrons. Both edges support low-lying excitations corresponding to
long-wave-length density fluctuations. However, for now we only consider the
ground state configuration. As a realistic model for the external potential
confining the electrons to the finite QH sample, we assume a background of
positive charges to be present that would exactly neutralize the electron
density if each lowest-Landau-level orbital were occupied with probability
$\nu$. As the local filling factor for the ground-state configuration depicted
in Fig.~\ref{phconj} deviates from its bulk value close to the edge, a dipolar
strip forms at the edge. While the system is not
charge-neutral {\em locally\/}, the 2D charge density integrated perpendicular
to the edge at any fixed location along the edge yields zero. We call this
weaker version of charge neutrality {\em 1D-local\/} charge neutrality. Note
that, in general, 1D-local neutrality will be violated when edge-density
fluctuations are present at the inner and/or outer edges.

Conceptionally, it is convenient to work consistently in the Hilbert space for
{\em holes\/} which has been truncated such that there are no states available
with guiding centers beyond the physical boundary of the sample. In hole
language, the ground-state configuration depicted in Fig.~\ref{phconj} is that
of a hole system which has undergone phase separation into two QH strips:
the inner one which is in the Laughlin state for filling factor $1-\nu=1/m$,
and an outer one extending from the physical sample boundary to the outer edge
which has filling factor one. In equilibrium, the energy $\mu_{\mathrm o}$ to
add a hole at the outer edge must equal the energy $\mu_{\mathrm i}$ to add
it at the inner edge: 
\begin{equation}\label{equcond}
\mu_{\mathrm o}=\mu_{\mathrm i}=\mu\quad ,
\end{equation}
where $\mu$ is the chemical potential. (Holes cannot be added at the physical
boundary of the sample due to the truncation of the Hilbert space.) To add a
hole at the inner edge, it must be provided with the energy per particle
$\zeta(1-\nu)$ of a locally charge-neutral hole system that is in the Laughlin
state for filling factor $1/m$, plus the electrostatic energy arising from the
added hole's interaction with the above-mentioned dipolar strip of
non-neutralized charges. A similar reasoning applies when adding a hole to the
outer edge. For $d>\ell$, we find~\cite{smalldnote}
\begin{eqnarray}
\mu_{\mathrm i} &=& \zeta(1-\nu) + \frac{e^2}{\epsilon\ell\pi}\,\,
\frac{d}{\ell}\,(1-\nu)\ln\left(\frac{1-\nu}{\nu}\right)\quad , \\
\mu_{\mathrm o} &=& \zeta(1) + \frac{e^2}{\epsilon\ell\pi}\,\,
\frac{d}{\ell}\,\ln\left(\frac{1}{\nu}\right)\quad,
\end{eqnarray}
where $\zeta(1)=-\sqrt{\pi/8}\,e^2/\epsilon\ell$ is the energy per particle
for a filled lowest Landau level. The equilibrium condition Eq.~(\ref{equcond})
then determines the ground-state edge separation $d$; it can be written as
\begin{equation}
\frac{\hbar v_{\mathrm J}}{\ell}\,\,\frac{d}{\ell} = \zeta(1-\nu) - \zeta(1)
\quad ,
\end{equation}
with a velocity
\begin{equation}\label{velJ}
v_{\mathrm J} = - \frac{e^2}{\epsilon\hbar\pi}\,\,\left[\nu\ln(\nu) + (1-\nu)
\ln(1-\nu)\right]\quad .
\end{equation}
Note that $\zeta(1-\nu)>\zeta(1)$,~\cite{morfnote} which means that we find a
positive value for $d$, as is required by self-consistency.~\cite{smalldnote}

\section{Energy of Long-Wave-Length Edge Excitations}

The inner edge of a QH sample that is at $\nu=1-1/m$ supports low-lying
excitations that correspond to fluctuations in the 1D edge
density which is obtained when integrating the 2D density profile of the inner
hole strip over the coordinate perpendicular to the edge. Similarly, the
outer hole strip has low-lying excitations that are density fluctuations
located at the outer edge. (Note that the edge of the outer hole strip that
coincides with the sample boundary originates from the truncation of the
Hilbert space in which we perform the particle-hole conjugation and does {\em
not\/} support physical excitations.) The local electric field, which is
related to the slope of the effective external potential felt by holes, has
opposite sign at the location of the inner and outer edges. Hence, density 
fluctuations at the inner and outer edges propagate in opposite direction.
Denoting 1D edge-density fluctuations
localized at the inner and outer edges by $\varrho_{\mathrm i}(x)$ and
$\varrho_{\mathrm o}(x)$, respectively, where $x$ is the coordinate along the
edge perimeter, we can write the energy of the system (measured from that of
the ground-state configuration depicted in Fig.~\ref{phconj}) as a functional
$E\big[\varrho_{\mathrm i}(x),\varrho_{\mathrm o}(x)\big]$. In the
long-wave-length limit, this functional can be expanded,
\begin{equation}\label{expand}
E\big[\varrho_{\mathrm i}(x),\varrho_{\mathrm o}(x)\big] = \sum_{\alpha,\beta
\in\{\mathrm i,o \}}\int\;\int dx\, dx^\prime \,\, \varrho_\alpha(x)\,
\varrho_\beta(x^\prime) \,\,\frac{\delta^2 E}{\delta\varrho_\alpha\,\delta
\varrho_\beta}(x, x^\prime) \quad + \quad \dots \quad ,
\end{equation}
yielding quadratic terms to lowest order in density fluctuations. Note that the
presence of edge excitations leads to non-neutralized charges on a length scale
given by the wave length of the edge-density fluctuation. As we consider the
case where holes interact via unscreened Coulomb interactions, the expansion
coefficients in Eq.~(\ref{expand}) diverge as the wave length of the
edge-density fluctuations increases, and become approximately equal at the
longest wave lengths which are of the order of the edge perimeter $L$. In that
limit, the edge-excitation normal modes are~\cite{wen} (i)~the charged
mode $\varrho_{\mathrm c}=\sqrt{1/\nu}\,\big(\varrho_{\mathrm o}+
\varrho_{\mathrm i}\big)$ which corresponds to fluctuations in the 
{\em total\/} edge charge density, and (ii)~the neutral mode
$\varrho_{\mathrm n} = \sqrt{(1-\nu)/\nu}\,\big[\varrho_{\mathrm o}+
\varrho_{\mathrm i}/(1-\nu)\big]$. The energy functional expressed in the
normal modes reads
\begin{equation}
E\big[\varrho_{\mathrm i}(x),\varrho_{\mathrm o}(x)\big] \equiv
E\big[\varrho_{\mathrm c}(x), \varrho_{\mathrm n}(x)\big]=
\frac{1}{2}\int d x \, dx^\prime \,\, V_{\mathrm c}(x-x^\prime)\,\,
\varrho_{\mathrm c}(x)\,\varrho_{\mathrm c}(x^\prime)+ \pi\hbar v_{\mathrm n}
\int d x\,\,\left[\varrho_{\mathrm n}(x)\right]^2\quad .
\end{equation}
\begin{figure}
\centerline{\psfig{figure=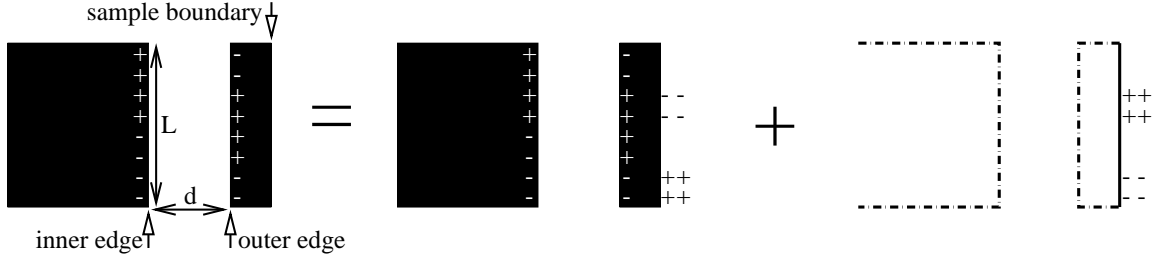,width=6in}}
\caption{Illustration of our procedure to separate long-range and short-range
pieces of interactions between edge-density waves. Depicted on the l.h.s.\ is
the real physical sample with edge-density waves present at
the inner and outer edges. Long-wave-length edge-density fluctuations give rise
to non-neutralized charges on a length scale that is of the order of the edge
perimeter $L$. The first term of the r.h.s.\ shows a fictitious system where
the positive background charge has been adjusted such that the charge density
integrated perpendicular to the edge yields zero {\em at any fixed location
along the edge\/}. Per definition, the fictitious system is always 1D-locally
neutral even when edge excitations are present, and long-range electrostatic
forces are cut off at distances larger than the edge width $d/\nu\ll L$. The
long-range Coulomb forces are accounted for in the second term on the r.h.s.\ 
where no charge imbalance occurs in the direction transverse to the edge.
\label{neutral}}
\end{figure}
The long range of Coulomb interactions enters only the dispersion of the
charged mode, which is actually the EMP mode predicted by classical
hydrodynamic theory.~\cite{volkmikh} The neutral mode propagates in the
direction opposite to the EMP mode; it has a linear dispersion with a velocity
$v_{\mathrm n}$ that depends on residual short-range interactions and is much
smaller than the characteristic velocity $\sim e^2/(\epsilon\hbar)\times \ln(L
/\ell)$ for EMP propagation. To actually determine $v_{\mathrm n}$, a careful
separation of long-range and short-range-interaction contributions to the
expansion coefficients in Eq.~(\ref{expand}) has to be performed. We were able
to achieve such a separation~\cite{uz:prb:98} by relating the energy of an
excitation in the physical QH sample to that of a fictitious system where the
background of positive charges has been adjusted to preserve 1D-local charge
neutrality, i.e, the 2D charge-density in the fictitious system integrated
perpendicular to the edge yields zero at any fixed location along the edge.
(See Fig.~\ref{neutral}.) Hence, the fictitious system looks neutral on length
scales that are larger than the edge width, and effective interactions are
short-ranged. The difference in energy in the real and fictitious systems 
accounts for the long-range electrostatic interactions and can be calculated
straightforwardly. Our approach to separate the long-range and short-range
parts of the interaction was inspired by a similar treatment~\cite{silin} of
long-range Coulomb interactions within Landau's Fermi-liquid theory. As the
central result of our calculations, we find the velocity of the
counterpropagating neutral mode,
\begin{equation}
v_{\mathrm n} = v_{\mathrm J}\quad,
\end{equation}
with $v_{\mathrm J}$ given by Eq.~(\ref{velJ}). We see that the velocity of the
neutral mode is directly related to the stiffness of the edge against
deviations from the ground-state value $d$ of the edge separation. Stability of
the two-branch edge and, therefore, consistency of the description based on
particle-hole conjugation requires a finite value of $v_{\mathrm n}$. Note
that the ratio of the characteristic EMP velocity to $v_{\mathrm n}$ is
approximately given by $\ln(L/\ell)$ which is $\sim 10$ in typical samples.

\section{Conclusion}

We have studied the two-branch edge realized in a quantum-Hall sample that is
at a filling factor $\nu=1-1/m$. A neutralizing background of positive charges
was assumed to confine the electrons to the finite sample, and long-range
Coulomb interactions have been taken into account. We find the separation $d$
between the inner and outer edges in the ground state as well as the stiffness
against deviations from $d$. This stiffness can be parameterized in terms of a
velocity $v_{\mathrm J}$ which turns out to be equal to the velocity of the
neutral counterpropagating edge mode.

\section*{Acknowledgments}

This work was supported in part by NSF grant Nos.\ DMR-9714055 (Indiana) and
DMR-9632141 (Florida) as well as Sonderforschungsbereich~195 of Deutsche
Forschungsgemeinschaft.


\begin{thebibliography}{10}
\bibitem{smalldnote}
Here and in the following, we neglect corrections which arise when $d\sim\ell$.
  Incorporation of these corrections is straightforward~\cite{uz:prb:98} and
  changes our results quantitatively by $\sim 10$\%.
\end{thebibliography}
\end{document}